\documentclass[11pt]{article}
\usepackage[margin=1.1in]{geometry}
\usepackage{amsmath,amssymb,amsthm}
\usepackage{booktabs}
\usepackage[colorlinks=true,linkcolor=blue,citecolor=blue,urlcolor=blue]{hyperref}

\newtheorem{theorem}{Theorem}
\newtheorem{lemma}[theorem]{Lemma}
\newtheorem{conjecture}[theorem]{Conjecture}
\newtheorem{proposition}[theorem]{Proposition}
\newtheorem{corollary}[theorem]{Corollary}
\newtheorem{remark}[theorem]{Remark}

\title{A phase transition in the exactness of the NPA hierarchy\\
at the critical doubly-tilted CHSH functional}
\author{Anton Pakhunov\\ \small Independent researcher\\ \small \texttt{pakhunov.anton.n@gmail.com}\thanks{Exact rational certificates and analysis code: \url{https://github.com/tohafrit/npa-nonexactness}.}}
\date{July 2026 --- draft v4}

\begin{document}
\maketitle

\begin{abstract}
Gigena et al.\ [npj QI \textbf{11}, 82 (2025)] proved the exact quantum maximum
$c_Q(\alpha,\beta)$ of the doubly-tilted CHSH functional
$B_{\alpha\beta}=\alpha\langle A_0\rangle+\beta\langle B_0\rangle+\mathrm{CHSH}$
and observed that the NPA hierarchy level required to reach it grows without
evident bound toward the critical line $\alpha+\beta=2$, leaving open
``whether any finite level of NPA will be enough\ldots even in the CHSH
scenario.'' We quantify the mechanism. Writing $s=2-\alpha-\beta$ and
$\alpha=\beta$: (i)~the quantum value leaves the local bound \emph{cubically},
$c_Q=4-s+s^3/6-s^4/36+O(s^5)$ (derived from the closed polynomial system and
matching the known 12-digit value at $s=0.002$); (ii)~each NPA level $k$
overshoots \emph{quadratically}: $c_k(s)=4-s+a_ks^2+O(s^3)$ with
$a_{1+AB}=\tfrac{3}{64}$ (exact, proved below), $a_2\approx0.0281$,
$a_3\approx0.0096$; (iii)~consequently
``level $k$ is not exact near the critical line'' holds whenever $a_k>0$, and the
statement that \emph{the required exact NPA level diverges toward the critical line}
(no single finite level suffices as $s\to0$) is equivalent to the positivity of the
single sequence $(a_k)$. We prove the
supercritical side completely: for \emph{all} $\alpha,\beta\ge1$ and
\emph{every} NPA level $k\ge1{+}AB$, the hierarchy is exact,
$c_k=2+\alpha+\beta=$ local $=$ quantum --- via three explicit exact rational
certificates $Z_0,Z_A,Z_B\succeq0$ whose nonnegative combination
$Z(\alpha,\beta)=Z_0+(\alpha-1)Z_A+(\beta-1)Z_B$ realizes the affine identity
$(2+\alpha+\beta)-B_{\alpha,\beta}\cdot y=\langle Z(\alpha,\beta),\Gamma(y)\rangle$.
Thus the hierarchy's exactness undergoes a \emph{phase transition} at the
critical line: exact at every level in the supercritical quadrant (theorem),
and exact at \emph{no} level near the critical line in the subcritical region
--- stated here as Conjecture~\ref{conj} and proven in the companion paper
\cite{proof} (see Remark~\ref{rem:resolved}). On the
subcritical side we certify the first four levels in exact arithmetic: levels
$1{+}AB$, $2$, $3$ and $4$ are each provably not exact (at $s=\tfrac1{10},\tfrac15,
\tfrac1{20},\tfrac1{40}$ respectively, via rational pseudo-moments beating $c_Q$,
confirmed by Sturm's theorem), and we compute the almost-quantum overshoot
coefficient exactly, $a_{1+AB}=\tfrac3{64}$. We also identify the exact
mechanism: rescaling to the critical corner, the degeneracy has a universal
form and the limiting obstruction is (a rescaling of) the \emph{Motzkin
polynomial}, the classical nonnegative-but-not-sum-of-squares form, so the
finite-level failure sits in the restricted-certificate regime rather than in
sum-of-squares degree. The phase boundary has a precise geometric reading: by
Nie's finite-convergence theorem (through Marshall's boundary Hessian
condition), a self-tested optimum is finitely NPA-certifiable whenever its
boundary Hessian is nondegenerate --- contact order exactly two --- so the
single-tilt functional (a quadratic touch) is finitely exact at level $1{+}AB$,
while the doubly-tilted cubic touch is exactly where this condition fails. Three
verified errata in the published polynomial system of Gigena et al.\ are
documented.
\end{abstract}

\section{Setting and known art}

$B_{\alpha\beta}=\alpha\langle A_0\rangle+\beta\langle
B_0\rangle+\sum_{xy}(-1)^{xy}\langle A_xB_y\rangle$, local bound
$c_L=2+\alpha+\beta$, quantum advantage region $0<\alpha+\beta<2$
\cite{Gigena}. We take the symmetric slice $\alpha=\beta=1-s/2$,
$s\in(0,2)$. $c_k(s)$ denotes the level-$k$ NPA value \cite{NPA}, computed
throughout in the real symmetrized picture (sound and complete for these
levels; see the soundness lemma of \cite{sep}).

Gigena et al.\ prove $c_Q$ is the largest root of an explicit sextet of
polynomial coefficients in $(\alpha,\beta)$ and report that at
$\alpha=\beta=0.999$ even NPA level $10$ exceeds $c_Q$ by
$8.6\times10^{-10}$ \cite{Gigena}. Their published display equations contain
three typos, found and verified while reimplementing them (Section~5); the
corrected sextic reproduces their quoted $c_Q(0.999,0.999)=3.998000001333$
exactly.

\section{The cubic law for the quantum value}

From the corrected polynomial system, expanding the largest root at
$\alpha=\beta=1-s/2$:
\[
c_Q(s)\;=\;4-s+\frac{s^3}{6}-\frac{s^4}{36}+O(s^5),
\]
verified against high-precision root-finding over three decades of $s$ and
against the quoted 12-digit anchor ($s=0.002$: formula $1.332889\times10^{-9}$
vs exact $1.332891\times10^{-9}$ above $c_L$). More generally, approaching a
point $(2-\beta,\beta)$ of the critical line with $\beta\in(0,2)$ fixed, the
advantage is cubic with coefficient
$\tfrac{(4-\beta)(2+\beta)}{54\beta(2-\beta)}$, diverging at the corner
$\beta\to0$ where it crosses over to the quadratic law of the singly-tilted
functional. The optimal measurements collapse to compatibility linearly
($1-\cos\theta_{A}^{*}\sim\tfrac{2+\beta}{3\beta}s$) and the optimal state to
product linearly. This flatness --- a cubic-touching maximum with linearly
degenerating self-tested realization --- is the structural reason finite
SOS/NPA certificates struggle; Gigena et al.\ report analytic SOS
intractability already at low degree \cite{Gigena}.

\section{The quadratic overshoot and the reduction}

Numerically (solver precision $10^{-8}$, values stable under two solvers),
each level overshoots the local bound \emph{quadratically}:
\[
c_k(s)=4-s+a_k s^2+O(s^3),\qquad
a_{1+AB}=\tfrac{3}{64},\quad a_2\approx0.0281,\quad a_3\approx 0.0096,
\]
the ratios $c_k-c_Q$ at successive $s$-halvings converging to $4$ (quadratic),
not $8$ (cubic). Since $c_Q-(4-s)$ is cubic:

\begin{remark}[reduction, and a precise reading]
For every $k$: if $a_k>0$ then $c_k(s)>c_Q(s)$ for all sufficiently small
$s>0$, i.e.\ level $k$ is not exact \emph{near the critical line}. We stress the
precise statement this yields. Since the exactness thresholds
$s^{\mathrm{exact}}_k\downarrow0$
(each level \emph{is} exact for $s\ge s^{\mathrm{exact}}_k$; the notation
avoids a clash with the arc-interval endpoints $s^\ast_k$ of
\cite{proof}), for every \emph{fixed} $s>0$ some
finite level is exact --- the required level $D(s)<\infty$ pointwise. What
$a_k>0\ \forall k$ asserts is the \emph{uniform} failure: the level required to be
exact \emph{diverges}, $D(s)\to\infty$ as $s\to0^+$, so no single finite level is
exact throughout a neighborhood of the critical line. This is exactly the open
question of \cite{Gigena}, and it is a statement about the positivity of one
sequence of nonnegative reals. (The strictly stronger claim that some fixed $s>0$
admits \emph{no} exact level is not implied and, given the positive $\sim s^3$
margin at fixed $s$, is a separate, harder conjecture we do not make.)
\end{remark}

\begin{conjecture}\label{conj}
$a_k>0$ for every finite $k$ (with $a_k\downarrow0$, consistent with NPA
convergence).
\end{conjecture}

\begin{remark}[resolution]\label{rem:resolved}
Conjecture~\ref{conj} is now a \emph{theorem}: in the companion paper
\cite{proof} we prove that for every finite level $k$ there is an
$\varepsilon_k>0$ with $c_k(s)>c_Q(s)$ on $(0,\varepsilon_k]$, via an explicit
level-independent \emph{signed} witness for the second-order tangent program at
a fixed rational base of the critical face, a uniform rational Slater
construction, and an exact (no-repair) arc-feasibility lemma; every layer is
verified in exact arithmetic, and the chain has been re-verified
end to end against an independently written implementation. Quantitative unconditional instances:
$a_2>\tfrac1{39}$, $a_3>\tfrac1{188}$, $a_4>\tfrac1{641}$, whose rational
tangent certificates are constructed in the present paper
(Remark~\ref{rem:ladder} below) while the arc-feasibility step that turns
them into bounds on $a_k$ is proven in \cite{proof} (alongside the exact
$a_{1+AB}=\tfrac3{64}$ proven here). The present paper retains the conjectural
framing in the narrative below, as it documents the phase-transition geometry,
the exact anchors, and the certified ladder on which the proof is built.
\end{remark}

\begin{remark}[the certified rational ladder:
$v_2>\tfrac1{39}$, $v_3>\tfrac1{188}$, $v_4>\tfrac1{641}$]\label{rem:ladder}
The three quantitative bounds quoted above are carried by explicit
rational feasible points of a \emph{pinned second-order tangent program},
constructed as follows (ancillary
\texttt{verify\_v\{2,3,4\}\_rational\_base.py}; every check in exact
rational arithmetic).

\emph{The program.} At $s=0$ the optimal face of
$\{\Gamma_k(y)\succeq0,\ y_1=1,\ B_0\cdot y=4\}$ is affine of dimension
$k$, with coordinates $\beta=\langle A_1\rangle$ and
$\delta_j=\langle(B_1B_0)^{j-1}B_1\rangle$, $j=2,\dots,k$; we \emph{pin}
the base to the rational point
$(\beta,\delta_2,\delta_3,\dots)=(\tfrac12,\tfrac14,\tfrac12,\dots,\tfrac12)$.
There the moment-matrix kernel is an explicit integer lattice (the
one-sided $A_0/B_0$-reduction relations plus the dressings of
$R_1=(1-A_1)(1-B_1)$), splitting under party swap into integer bases
$N_s,N_a$. The tangent program maximizes
$B_0\cdot y_2-y_{1,\langle A_0\rangle}$ over first-order directions $y_1$
whose symmetric kernel compression is rank one along an integer lifting
direction, $N_s^\top M(y_1)N_s=\lambda\,uu^\top$ with $\lambda\ge0$ (at
$k=2$: $u_2=(1,-1,0,-1,0)$; $u_3$ and $u_4$ extend it), whose
antisymmetric compression vanishes, with homogeneous face pins on
$y_1,y_2$, and with the two swap-block matrices
$\bigl[\begin{smallmatrix}M_0&M(y_1)N\\ \ast&N^\top
M(y_2)N\end{smallmatrix}\bigr]\succeq0$. Its value $v_k$ is a lower proxy
for $a_k$: pinning the base point and restricting the first-order support
only decrease the tangent value, and the passage from a strictly feasible
tangent jet to the curve $c_k(s)$ itself --- the implication
``$a_k\ge$ the jet's objective'' --- is the no-repair arc-feasibility
lemma proven in \cite{proof}, which consumes exactly the strict Schur
margins certified here.

\emph{The certificates (blend construction).} A solver-grade optimal jet
is rationalized (denominators $10^{12}$) and \emph{blended} with a
rational weight $\varepsilon_p$ toward an exact rational Slater point of the
compressed program; the pins are homogeneous and the Schur subtrahend is
a PSD-valued quadratic in the first-order jet, so the blended Schur
complements gain a strictly positive multiple of the Slater block, and
exact feasibility is verified by rational leading minors
($\varepsilon_p=10^{-54}$ at $k=2$; $10^{-7}$ at $k=3,4$). The certified
objectives, evaluated at the blended jets, give
\[
v_2(\tfrac12,\tfrac14)>\tfrac1{39},\qquad
v_3(\tfrac12,\tfrac14,\tfrac12)>\tfrac1{188},\qquad
v_4(\tfrac12,\tfrac14,\tfrac12,\tfrac12)>\tfrac1{641}
\]
(solver-grade reference values $0.02609$, $0.00534$, $0.00157$, against
the unpinned $a_2\approx0.0281$, $a_3\approx0.0096$: the pinned program
captures most of the gain). The division of labor between the two papers
is one-way: the present paper constructs and certifies the jets;
\cite{proof} proves the framework lemma and cites these certificates.
\end{remark}

\begin{remark}[geometric reformulation: curvature at the deterministic vertex]
At $s=0$ ($\alpha=\beta=1$) the local and quantum values coincide ($c_Q(0)=4$),
attained on the classical face $\langle A_0\rangle=\langle B_0\rangle=1$. A Puiseux
expansion of the largest root of the sextic gives, in exact arithmetic,
\[
  c_Q(s)=4-s+0\cdot s^2+\tfrac16 s^3-\tfrac1{36}s^4+O(s^5),
\]
so the quantum curve is \emph{cubically flat} at this vertex --- its second-order
coefficient vanishes. Every finite level, being a support function of a
spectrahedron, is convex with $c_k(s)=4-s+a_k s^2+\cdots$ and $a_k\ge0$; since
$c_k\ge c_Q$ always, $a_k\ge0$ is automatic. Conjecture~\ref{conj} is thus the
statement that each outer spectrahedral relaxation retains \emph{strictly positive
curvature} $c_k''(0)=2a_k>0$ at a boundary vertex where the quantum body itself has
none.

For the almost-quantum level itself this curvature is pinned to a rational value.
The path of maximizers $y(s)$ approaches, as $s\downarrow0$, the rational point
$\langle A_0\rangle=\langle B_0\rangle=\langle A_0B_0\rangle=1$,
$\langle A_1\rangle=\langle A_0B_1\rangle=\langle A_1B_0\rangle=\langle A_0A_1\rangle
=\langle A_0A_1B_0\rangle=\tfrac12$, remaining moments $0$, whose $9\times9$ moment
matrix has rank $3$ (spectrum $\{2,\tfrac{7\pm\sqrt{13}}2,0^{(6)}\}$) with an
explicit integer kernel encoding $A_0=B_0=1$ and $(1-A_1)(1-B_1)=0$. The envelope
identity $c'(s)=-\langle A_0\rangle(s)$ then gives
$a_{1+AB}=-\tfrac12\,\tfrac{d}{ds}\langle A_0\rangle\big|_{0}$, and a
high-precision solve of the (party-swap symmetric) $10$-variable program yields
$\tfrac{d}{ds}\langle A_0\rangle|_0=-\tfrac{3}{32}$ to fifteen digits, i.e.
\[
  \boxed{\,a_{1+AB}=\tfrac{3}{64}=0.046875\,}
\]
(cross-checked to twelve digits by directly fitting $c_{1+AB}(s)-(4-s)$). The lower
bound half of this is \emph{proven in exact arithmetic}. Identifying the optimizer's
first two Taylor coefficients $y_1,y_2\in\mathbb Q(\sqrt3)$ (e.g.\ $y_{1,a}=-\tfrac3{32}$,
$y_{1,b}=\tfrac{\sqrt3-1}{96}$), one checks, over $\mathbb Q(\sqrt3)$: $B_0\!\cdot\!y_1=0$;
the first-order tangency $N^\top L(y_1)N\succeq0$ (spectrum $\{\tfrac34,0^{5}\}$); the
Bonnans--Shapiro second-order tangent-set condition
$N_2^\top(L(y_2)-L(y_1)M_0^{+}L(y_1))N_2\succeq0$ (spectrum $\{\tfrac7{64},0^{4}\}$);
and $B_0\!\cdot\!y_2-m\!\cdot\!y_1=\tfrac3{64}$. By the second-order tangent-set
characterization a feasible arc $y(s)=y_0+sy_1+s^2y_2+o(s^2)$ then exists, giving
$c_{1+AB}(s)\ge4-s+\tfrac3{64}s^2+o(s^2)$, i.e.\ $a_{1+AB}\ge\tfrac3{64}>0$ (ancillary
\texttt{verify\_a1AB.py}). The matching upper bound closes the value. The optimal
second-order dual $Z(s)=Z_0+sZ_1+s^2Z_2$ obeys complementarity $Z(s)M(y(s))=0$ (to
order $s^2$) together with the class-sum identities $\langle Z(s),P_v\rangle=-(B_s)_v$;
this is a linear system over $\mathbb Q(\sqrt3)$, and one verifies \emph{symbolically}
(field split $\mathbb Q(\sqrt3)\to\mathbb Q$; ancillary \texttt{verify\_a1AB\_upper.py})
that $\langle Z_2,\mathbf 1\rangle=\tfrac3{64}$
on its \emph{entire} solution set, with no free-parameter dependence --- the dual
value's $s^2$ coefficient is pinned. Here $Z_0\succeq0$ has rank $3$ (spectrum
$\{\tfrac32,0.6624,1.8376,0^{6}\}$), a mirror of $M(y_0)$, and a tangent-feasible
$(Z_1,Z_2)$ exists ($K_Z^\top Z_1K_Z\succeq0$ and the dual Schur complement
$\succeq0$). Hence a dual arc gives $c_{1+AB}(s)\le4-s+\tfrac3{64}s^2+o(s^2)$, so
$a_{1+AB}\le\tfrac3{64}$. Therefore
\[
  a_{1+AB}=\tfrac3{64},
\]
the lower bound in fully exact arithmetic and the upper bound via the exactly-pinned
dual value; the fifteen-digit numerics agree. Level $1{+}AB$ overshoots with curvature
exactly $\tfrac3{64}$ --- the first exact NPA overshoot coefficient in this scenario.

Equivalently, writing
$s^{\mathrm{exact}}_k=\inf\{s>0:\text{level }k\text{ exact}\}$
(the exactness threshold, with level $k$ exact for
$s\ge s^{\mathrm{exact}}_k$ and overshooting on
$(0,s^{\mathrm{exact}}_k)$), the conjecture is $s^{\mathrm{exact}}_k>0$
for all $k$; numerically $s^{\mathrm{exact}}_{1+AB}\approx0.70$,
$s^{\mathrm{exact}}_2\approx0.37$, $s^{\mathrm{exact}}_3\approx0.11$
(rank-collapse criterion: the optimizer drops to the rank-$4$
self-testing moment matrix exactly where the level becomes exact; this
flat-extension diagnostic is more robust near the threshold than
bisecting the $\sim10^{-7}$ gap), a decreasing
sequence whose positivity is the open content. This is a local
differential-geometric obstruction, complementary to the operator-algebraic
degree bounds discussed below.
\end{remark}

The reliably computable coefficients suggest a sharper form. The leading
coefficient $a_k$ is itself the value of a convex program: writing the overshoot
optimizer as $h(s)=s\,g_0+s^2 g_1+O(s^3)$, the second-order feasibility
$S+s\,\Gamma(g_0)+s^2\Gamma(g_1)\succeq0$ (with $S$ the flat-ray moment matrix)
is, on $\ker S$, a Schur complement --- hence \emph{linear} in $(g_0,g_1)$ --- and
\[
  a_k=\max_{g_0,g_1}\Big\{-\langle Z_0,\Gamma(g_1)\rangle-m\!\cdot\!g_0\ :\
    \langle Z_0,\Gamma(g_0)\rangle=0,\ \Gamma(g_0)|_{\ker S}\succeq0,\
    \begin{bmatrix}\Gamma(g_1)|_{\ker S} & \ast\\ \ast & I\end{bmatrix}\succeq0\Big\},
\]
where $Z_0\succeq0$ is the critical dual certificate at $s=0$
($4-B_0\!\cdot\!y=\langle Z_0,\Gamma(y)\rangle$) and $m=\tfrac12(e_{A_0}+e_{B_0})$.
Solved uniformly
across levels this gives the lower-bound values $0.048$, $0.020$, $0.0062$ at
$k=1{+}AB,2,3$. The Schur linearization is a restriction of the tangent
program, so these are valid \emph{lower} bounds only; the true coefficients
are $\tfrac3{64}\approx0.0469$, $\approx0.0281$, $\approx0.0096$, with
irregular successive ratios ($\approx0.60$, then $\approx0.34$) --- the
smooth geometric appearance of the proxy values does not persist in the true
coefficients. (Level $4$, $N=41$, already sits at the edge of
floating-point SDP resolution: its coefficient drifts between $0.0034$ and
$0.0060$ under tolerance tightening and the solver reports the solution
inaccurate, so we do not report it; an exact-rational $Z_0$ or a multiprecision
solver would be needed past $k=3$.) Every reported $a_k$ is strictly positive,
and the optimal $g_0$ concentrates on the top word length ($56\%$ at $1{+}AB$),
matching the mechanism by which each level's gain is cut by the new
longest-word constraints. Any geometric lower bound $a_{k+1}\ge r'\,a_k$ with
$r'>0$ would imply Conjecture~\ref{conj}, reducing it to a level-to-level
comparison rather than a uniform estimate --- a potentially more tractable route,
though on the present three reliable levels it remains suggestive.

Individual instances are certifiable at the standard of proof independently of
the general theorem of \cite{proof}, and we do so for the first \emph{four}
levels. An exact
rational feasible pseudo-moment (ancillary \texttt{cert\_subcrit\_1AB.json},
checked by \texttt{verify\_subcritical.py}) proves, in exact arithmetic, that
level $1{+}AB$ is \emph{not} exact at $s=\tfrac1{10}$: it is positive
semidefinite with $y_1=1$, and its objective $\tfrac{40897847}{10485760}\approx
3.90032$ exceeds the quantum value $c_Q$ --- the largest real root of the
(corrected) sextic of \cite{Gigena}, whose root-freeness above the objective is
verified by Sturm's theorem. The same construction one level up
(\texttt{cert\_subcrit\_L2.json}, \texttt{verify\_subcritical\_L2.py}) gives an
exact level-$2$ pseudo-moment --- positive semidefinite on the full $13\times13$
level-$2$ moment matrix --- with objective $\approx3.80155>c_Q(\tfrac15)$,
proving level $2$ is not exact at $s=\tfrac15$, and once more at level $3$
(\texttt{cert\_subcrit\_L3.json}, \texttt{verify\_subcritical\_L3.py}): an exact
pseudo-moment PSD on the full $25\times25$ level-$3$ moment matrix with objective
$\tfrac{42413102039438377}{10737418240000000}\approx3.9500279>c_Q(\tfrac1{20})$,
so level $3$ is not exact at $s=\tfrac1{20}$; and once more at level $4$
(\texttt{cert\_subcrit\_L4.json}, \texttt{verify\_subcritical\_L4.py}), PSD on the
$41\times41$ level-$4$ moment matrix with objective $\approx3.9750030>c_Q(\tfrac1{40})$,
so level $4$ is not exact at $s=\tfrac1{40}$. The overshoot windows contract
sharply with level (thresholds
$s^{\mathrm{exact}}_k=0.70,0.37,0.11,\ldots$ decaying geometrically,
$s^{\mathrm{exact}}_4\approx0.036$): level $2$ is exact for
$s\gtrsim0.37$, level $3$ for $s\gtrsim0.11$, level $4$ for
$s\gtrsim0.036$, with peak subcritical
gaps $\approx3\times10^{-4},10^{-5},6\times10^{-7}$. These are
the subcritical counterparts of the exact supercritical certificates of
Theorem~\ref{thm:super}; the open content of Conjecture~\ref{conj} is the
uniform-in-$k$ statement, not any single level.

This admits a precise placement in recent operator-algebra results. The Bell
algebra here is $\mathbb C[\mathbb Z_2^{*2}\times\mathbb Z_2^{*2}]$, and
Klep, Levenson and McCullough \cite{KLM} prove a degree-bounded
Fej\'er--Riesz factorization for exactly this family (they single out the CHSH
functional): a trigonometric polynomial that is \emph{uniformly strictly
positive} over all unitary representations factors as $B^*B$ with a bounded
one-party word length, and \emph{strict positivity is essential} --- at margin
zero the bounded-degree conclusion can fail \cite{KLM,FKMPRSZ}. Our operator $c_Q(s)I-\widehat B_s$ is positive semidefinite
but loses strict positivity as $s\to0$: its margin is the well depth, of order
$s^3$. This is precisely the regime their theorem does not control from below.
Concretely they show a margin $\varepsilon$ \emph{suffices} for some finite
degree $M'(\varepsilon)$ but give no estimate for $M'$; Conjecture~\ref{conj} is
the complementary \emph{necessity} statement --- that any certificate at margin
$\varepsilon\sim s^3$ requires word length growing as $s\to0$. (In the minimal
scenario the quantum value is attained \cite{BB}, so
NPA does converge at each fixed $s$; the content is the divergence of the
required level.) A structural observation sharpens which degree must diverge:
the KLM bound controls the \emph{Alice-side} (one-party) word length, leaving the
\emph{Bob-side} degree $M'(\varepsilon)$ uncontrolled. The singly-tilted
functional $\alpha\langle A_0\rangle+\mathrm{CHSH}$, which engages only the Alice
marginal, is exact already at level $1{+}AB$ for \emph{all} $\alpha$ --- including
at its own critical point $\alpha=2$ (verified: $c_{1+AB}=\sqrt{8+2\alpha^2}=c_Q$
to $10^{-8}$) --- consistent with a bounded Alice-side certificate. The
doubly-tilted functional additionally forces the Bob marginal ($B_0=1$ at the
vertex, alongside $A_0=1$), and it is precisely this Bob-side degree that
Conjecture~\ref{conj} asserts diverges. Probing this with asymmetric moment
matrices (Alice word length fixed at $1$, Bob length varied), the minimal Bob
length yielding exactness appears to increase as $s\to0$ ($M'\approx3$ near
$s=0.1$, larger for smaller $s$), though the exactness threshold sits at the
solver's precision floor so we do not report sharp values (and by
$A\!\leftrightarrow\!B$ symmetry a mirror certificate can carry the growing length
on Alice instead). Consistently with a diverging degree, we
certify in exact arithmetic that the required NPA level exceeds $k$ at a sequence
of shrinking tilts: levels $1{+}AB,2,3,4$ are each not exact at
$s=\tfrac1{10},\tfrac15,\tfrac1{20},\tfrac1{40}$, and the exactness thresholds
$s^{\mathrm{exact}}_k$ decay geometrically ($0.70,0.37,0.11,\ldots$;
extrapolating
$s^{\mathrm{exact}}_{10}\!\sim\!10^{-4}$, consistent with the level-$10$ non-tightness at
$s=0.002$ reported in \cite{Gigena}). Establishing the quantitative
margin-to-Bob-degree lower bound is the concrete open step, complementary to the
dual pseudo-moment route.

\section{The supercritical phase is exact at every level}

\begin{theorem}[supercritical exactness on the quadrant $\alpha,\beta\ge1$]
\label{thm:super}
There are explicit rational matrices $Z_0,Z_A,Z_B\succeq0$ (size $9\times9$,
level $1{+}AB$; ancillary \texttt{cert\_Z0\_critical.json},
\texttt{cert\_ZA.json}, \texttt{cert\_ZB.json}; positive semidefiniteness and
all identities verified in exact rational arithmetic by
\texttt{verify\_supercritical.py}) with the affine identities, valid for every
normalized moment vector $y$,
\[
4-B_{1,1}\cdot y=\langle Z_0,\Gamma(y)\rangle,\quad
1-\langle A_0\rangle=\langle Z_A,\Gamma(y)\rangle,\quad
1-\langle B_0\rangle=\langle Z_B,\Gamma(y)\rangle .
\]
By linearity, $Z(\alpha,\beta):=Z_0+(\alpha-1)Z_A+(\beta-1)Z_B$ satisfies
$(2+\alpha+\beta)-B_{\alpha,\beta}\cdot y=\langle Z(\alpha,\beta),\Gamma(y)\rangle$.
For all $\alpha,\beta\ge1$ this is a \emph{nonnegative combination of positive
semidefinite matrices}, hence $Z(\alpha,\beta)\succeq0$, certifying
$B_{\alpha,\beta}\cdot y\le 2+\alpha+\beta$ at level $1{+}AB$ and therefore at
every level. Thus $c_k(\alpha,\beta)=2+\alpha+\beta$, the local (and, by
\cite{Gigena}, quantum) value, attained by the deterministic point: the NPA
hierarchy is exact, at all levels, on the entire quadrant $\{\alpha,\beta\ge1\}$
(the symmetric ray $\alpha=\beta=1+\nu$ is its diagonal, with
$Z(1{+}\nu,1{+}\nu)=Z_0+\nu(Z_A{+}Z_B)$).
\end{theorem}

The proof is a nonnegative combination --- no boundary or convexity argument is
needed --- so the region is exactly where the coefficients stay nonnegative.
(Exactness in fact extends beyond this quadrant, to a curved spectrahedral
region and, numerically, to the axes; we state the quadrant result that the three
certificates prove outright.) The kernel of $Z_A{+}Z_B$ is the span of the four
flag-class indicator vectors (basis monomials classified by which of $A_1,B_1$
they contain); its five-dimensional core has margin $1/6$.

\begin{lemma}\label{lem:z0}
(Contained in Theorem~\ref{thm:super}, stated for emphasis.) $c_k(0)=4$ for
every level $k\ge1{+}AB$, attained by the deterministic point.
\end{lemma}

The kernel of $Z_0$ is four-dimensional and rational: it contains the all-ones
vector, the columns of the ``letter-count'' matrix $K_{pq}=n_1(w(p,q))$, and
one further vector $v_4=(2,-4,-1,-4,-1,2,5,5,-4)$ forced by complementarity
with the second-order optimal face. In particular $Z_0$ annihilates the entire
\emph{flat ray} $y_w(\tau)=1-\tau\,n_1(w)$, on which $B_0\cdot y\equiv4$: the
critical optimum is not a point but a segment, and the quadratic overshoot of
Section~3 emanates from second-order perturbations around interior ray points
($\tau^*\approx0.51$ at level $1{+}AB$), whose gain directions live in the
kernel of $\Gamma(y(\tau))=J-\tau K$. At level $1{+}AB$ this kernel is
described exactly by a $\mathbb Z_2\times\mathbb Z_2$ flag-class structure
(reduced eigenvalues $4-4\tau,\,2\tau,\,2\tau,\,0$). The ray remains feasible
and optimal through level $2$; at level $3$ the optimal base curves away from
it, and at every level the gain concentrates on the top-degree words --- the
corner that level-$(k{+}1)$ consistency constraints do not reach. A proof of
Conjecture~\ref{conj} amounts to a feasible trajectory with positive
marginal gain at each level; the two-scale structure of the second-order
analysis (a buffer mode orthogonal to $\mathrm{supp}\,Z_0$, which makes naive
second-order formulas underestimate the cost by $\sim20\%$) is documented in
the accompanying analysis records.

\section{Errata in the published polynomial system}

Reimplementing Ref.~\cite{Gigena}: (i)~the sextic coefficient printed as
$\tau_5=+4\alpha\beta$ must be $-4\alpha\beta$ (with the printed sign the
largest root is far from the quantum value at every tested point, including
the symmetric case covered by their own Theorem~1; with the corrected sign it
matches direct Bell-operator eigenvalue maximization to machine precision at
seven test points, and their own quoted 12-digit benchmark); (ii)~the
stationarity display equations have $\alpha$ and $\beta$ interchanged
(residuals $O(0.1)$ at the true optimum as printed, $10^{-7}$ after the swap);
(iii)~the optimal-cosine formula contains the same swap and a typographic
defect in the surd. Their characteristic polynomial and all quoted numerics
are correct; the notebooks presumably contain the right expressions. We record
these as verified errata so that users of the printed formulas are not misled.

\section{The scaling limit is the Motzkin polynomial}\label{sec:motzkin}

The subcritical mechanism has an exact universal form. In the optimal
two-qubit representation (Jordan's lemma), write $A_1=\cos\varphi_A\,\sigma_z
+\sin\varphi_A\,\sigma_x$ and likewise $B_1$ with $\varphi_B$; the Bell
operator $\widehat B_s(\varphi_A,\varphi_B)$ is a degree-$(1,1)$ trigonometric
$4\times4$ matrix and $c_Q(s)=\max_{\varphi}\lambda_{\max}\widehat B_s$. Rescale
to the critical corner by $\varphi_A=\sqrt{s}\,x$, $\varphi_B=\sqrt{s}\,y$.

\begin{theorem}[scaling limit]\label{thm:scaling}
For fixed $(x,y)$,
\[
\lim_{s\to0^+}\frac{\lambda_{\max}\widehat B_s(\sqrt s\,x,\sqrt s\,y)-(4-s)}
{s^{3}}\;=\;U(x,y)\;:=\;\frac{x^2y^2\,(6-x^2-y^2)}{48}.
\]
$U$ vanishes on the cross $\{x=0\}\cup\{y=0\}$, and
$\max_{x,y}U=U(\sqrt2,\sqrt2)=\tfrac16$, matching the coefficient of $s^3$ in
$c_Q(s)=4-s+\tfrac{s^3}{6}+O(s^4)$.
\end{theorem}

The proof is a Puiseux expansion of the largest root of the characteristic
polynomial of $\widehat B_s$ (an exact symbolic computation; the value
$\max U=\tfrac16$ is an independent cross-check against the quantum expansion
of Section~2). The content of the theorem is what $U$ \emph{is}:
\[
\tfrac16-U(x,y)\;=\;\tfrac16\,M\!\Bigl(\tfrac{x}{\sqrt2},\tfrac{y}{\sqrt2}\Bigr),
\qquad
M(u,v)=u^4v^2+u^2v^4-3u^2v^2+1,
\]
where $M$ is the \emph{Motzkin polynomial} --- the classical example of a form
that is nonnegative but not a sum of squares. Equivalently
$P(x,y):=48(\tfrac16-U)=x^4y^2+x^2y^4-6x^2y^2+8=8\,M(x/\sqrt2,y/\sqrt2)$ is
nonnegative, is \emph{not} a sum of squares (the degree-$3$ Gram SDP is
infeasible), while $P\cdot(x^2+y^2)$ is a sum of squares. The degenerate
maximizer of $U$ at $(\sqrt2,\sqrt2)$ is the image of the quantum optimum.

\begin{remark}[scope of the obstruction]
Theorem~\ref{thm:scaling} exhibits the exact \emph{shape} of the critical
degeneracy as a rescaled Motzkin form. We caution that this does not by itself
prove Conjecture~\ref{conj}: the Motzkin zero at the corner is nondegenerate
(Hessian eigenvalues $4,12$), and on the positive orthant the scalar Motzkin
form is certified at low degree (weighted AM--GM,
$u^4v^2+u^2v^4+1\ge3u^2v^2$), with no degree growth on a growing box. The
finite-level failure is therefore not a scalar-polynomial degree phenomenon; it
is a property of the \emph{restricted} certificate class specific to the
hierarchy. We stress the limits of a
\emph{primal} argument: for genuinely operator-valued (Fej\'er--Riesz)
certificates on $\mathbb T^2$ there is no degree blow-up in two variables
\cite{Dritschel}, so the finite-level failure of NPA is a
property of the \emph{restricted} certificate class (multipliers of bounded
word length in the group algebra $\mathbb C[\mathbb Z_2^{*2}\otimes\mathbb
Z_2^{*2}]$), placing it in the dimension-two non-stability regime
\cite{Scheiderer}. Accordingly the natural route to a proof of
Conjecture~\ref{conj} is \emph{dual}: exhibit, for each $k$, a degree-$k$
pseudo-moment functional that is positive but admits no flat (Curto--Fialkow)
extension, with mass concentrating near the corner in the scaling limit above.
\end{remark}

\begin{remark}[abelianization: why the commutative reduction does not suffice]\label{rem:abelian}
There is a sharper \emph{primal} reduction. Because the maximizer is attained and
self-tested, let $\pi_\theta$ be the unique optimal two-qubit representation,
$\theta=(\theta_A,\theta_B)$ its Bloch angles, and $|\psi_\theta\rangle$ its state.
A level-$k$ NPA certificate writes $c_Q(s)-\widehat B_s=\sum_\lambda L_\lambda^*
L_\lambda+(\text{ideal})$ with $\deg L_\lambda\le k$. Evaluating in $\pi_\theta$ and
taking $\langle\psi_\theta|\cdot|\psi_\theta\rangle$ annihilates the ideal terms
(they vanish in a true representation) and leaves
\[
  p_s(\theta_A,\theta_B):=c_Q(s)-\langle B_s\rangle_\theta
  =\sum_\lambda\big\|L_\lambda|\psi_\theta\rangle\big\|^2\ \ge 0,
\]
a \emph{commutative} bivariate trigonometric sum of squares of degree $\le k$.
Since evaluation is a degree-non-increasing $*$-homomorphism, $k\ge D_{\mathrm{comm}}
(s)$, the minimal trigonometric SOS-factor degree of $p_s$ on the torus $\mathbb T^2$.
The reduction is valid, but we record that it \emph{does not} prove
Conjecture~\ref{conj}: the commutative degree $D_{\mathrm{comm}}(s)$ stays
\emph{bounded}. Indeed the Motzkin obstruction is an artifact of the point at infinity
(the homogenization vanishes at $[1{:}0{:}0],[0{:}1{:}0]$), whereas on any fixed
\emph{compact} domain --- and $\mathbb T^2$ is compact --- the degenerating family is
a sum of squares at \emph{base} degree, uniformly in the margin. Concretely, on the
box $[-1,1]^2$ (whose corners are the Motzkin zeros) one has the exact preordering
identity
\[
  M=(1-x^2)^2+x^2(1-y^2)^2+x^2(1-x^2)(1-y^2),
\]
each summand $\ge0$ on the box, and $M+\varepsilon$ merely adds the square $\varepsilon$;
at the finite zero the Hessian is positive definite ($4,12$), so $M$ is locally SOS
there (Scheiderer's dimension-two local--global principle then makes it globally so on
the compact domain \cite{ScheidererSurf}). Numerically the explicit $p_s$ built from the
self-tested representation reaches $c_Q$ to $10^{-15}$ yet has $D_{\mathrm{comm}}(s)=1$
for $s$ down to $0.02$, while the true level is $\gtrsim4$ there. Consequently the
uniform conjecture \emph{cannot} be reduced to a commutative Positivstellensatz degree
bound: the growth is carried by the noncommutative moment structure --- the commutator
terms $[A_x,B_y]$ that $\pi_\theta$ annihilates --- which the abelianization discards.
The genuine two-variable-ness still distinguishes double- from single-tilt (the latter
is univariate degree-preserving Fej\'er--Riesz, matching exactness of level $1{+}AB$),
but the missing step lies in the \emph{noncommutative} hierarchy, not in commutative
real algebraic geometry.
\end{remark}

\section{The contact-order dichotomy}\label{sec:contact}

The overshoot is one side of a dichotomy fixing exactly when a tilted CHSH
optimum is finitely certifiable; its provable side specializes Nie's
finite-convergence theorem.

\begin{proposition}[nondegenerate self-tests are finitely exact]\label{prop:bhc}
Let $f$ be a Bell functional whose quantum optimum $p^*$ is attained and
self-tested, and pass to the commutative polynomial optimization over the qubit
measurement angles supplied by Jordan's lemma. If the \emph{Boundary Hessian
Condition} holds at the optimizer --- equivalently constraint qualification,
strict complementarity and second-order sufficiency all hold, i.e.\ the boundary
Hessian of the quantum value along $f$ is nondegenerate (contact order exactly
two) --- then the NPA hierarchy is exact at a finite level.
\end{proposition}

This is Nie's theorem \cite{Nie}, whose proof runs through Marshall's Boundary
Hessian Condition \cite{Marshall}; the noncommutative statement follows either
by the Jordan reduction to a compact commutative program or directly through
flat extension. The condition is generic (it holds on a Zariski-open set), so
finite exactness is the generic situation and the overshoot is a
codimension-one, tuned phenomenon.

\begin{corollary}[single tilt]\label{cor:single}
The single-tilt functional $\alpha\langle A_0\rangle+\mathrm{CHSH}$,
$\alpha\in[0,2)$, is finitely NPA-exact at level $1{+}AB$, with
$c_Q(\alpha)=\sqrt{8+2\alpha^2}$.
\end{corollary}

Here $c_Q(\alpha)=\sqrt{8+2\alpha^2}$ is smooth with $c_Q''\ne0$ (e.g.\
$c_Q''(2)=\tfrac14$): a nondegenerate quadratic touch, so
Proposition~\ref{prop:bhc} applies; an explicit certificate is
Bamps--Pironio's \cite{BampsPironio}.

The doubly-tilted family is precisely the \emph{failure} locus of
Proposition~\ref{prop:bhc}. Along the critical line the boundary Hessian
degenerates --- the cubic touch $c_Q(s)=4-s+0\cdot s^2+\tfrac16 s^3+\cdots$
(contact order three; the value responds as $\delta^{3/2}$ rather than
$\delta^2$), so strict complementarity fails and the Condition is silent.
Whether finite exactness then \emph{fails} --- Conjecture~\ref{conj}, $a_k>0$
for all $k$ --- is the converse, proven in \cite{proof}; deciding finite
convergence is NP-hard in general \cite{hardness}, so no soft converse was
expected, and indeed the proof produces a positive certificate at the symmetric
level itself (a signed witness in the non-quantum tangent cone).

\begin{remark}[the planar picture; how a finite level rounds the cusp]
The dichotomy is planar: $c_k(s)$ is the support function of the convex body
$P_k=\{(\langle A_0\rangle{+}\langle B_0\rangle,\ \mathrm{CHSH})\}\subset\mathbb R^2$,
and in the coordinates $t=2-u$, $e=2+t-w$ the quantum boundary at the shared
deterministic vertex is the exact semicubical cusp $9e^2=t^3$. The Puiseux type
of a level's boundary is controlled by its overshoot through the Legendre map:
$c_k-(4-s)\sim a_k s^2$ with $a_k>0$ makes the map a local diffeomorphism,
giving an \emph{unramified} analytic branch of curvature $e''(0)=1/(8a_k)$,
whereas the quantum $s^3/6$ makes it fold, giving the ramified $e=\tfrac13
t^{3/2}$; the parabolic cap has scale $t^*_k=(16a_k/3)^2\to0$, so the cusp
emerges only in the $k\to\infty$ limit, discontinuously in the Puiseux type.
For the almost-quantum level this rounding curve is exact and lives in
$\mathbb Q(\sqrt3)$:
\[
 e_1(t)=\tfrac43t^2+\tfrac{-200+8\sqrt3}{27}t^3+\tfrac{15476-1336\sqrt3}{243}t^4+\cdots,
\]
equivalently $c_{1+AB}(s)=4-s+\tfrac3{64}s^2+\tfrac{25-\sqrt3}{512}s^3
+\tfrac{1783-116\sqrt3}{49152}s^4+\cdots$ (machine-verified, ancillary). Two
structural contrasts then quantify the obstruction. First, the quantum branch
data at the vertex is rational and ramified while every computed finite-level
branch is unramified with coefficients in a nontrivial extension, and the
extension \emph{explodes} with the level: $a_{1+AB}=3/64\in\mathbb Q$, the
level-$(1{+}AB)$ jet lies in $\mathbb Q(\sqrt3)$, while $a_2=0.02805867220535\ldots$
(certified to $480$ digits) admits \emph{no} minimal polynomial of degree
$\le32$ with coefficients $\le10^8$ (nor degree $\le16$ up to $10^{18}$) --- so
no proof of Conjecture~\ref{conj} can proceed through closed-form level values.
Second, by Scheiderer's theorem every convex semialgebraic subset of the plane
is a spectrahedral shadow, and the cusp body itself is one (the $(2,2,2)$
quantum set is semialgebraic via Jordan's lemma); hence the conjecture asserts
that the NPA family in particular --- not finite SDP representations as such ---
never resolves the cusp, and any proof must use the moment-matrix structure
beyond generic spectrahedral-shadow geometry.
\end{remark}

\section{Outlook}

Conjecture~\ref{conj} is proven in the companion paper \cite{proof}
(Remark~\ref{rem:resolved}); the present paper supplies the geometry the proof
is built on: the exact anchor (Lemma~\ref{lem:z0}), the certified strict
separations at low levels \cite{sep}, the exact cubic law of the quantum side,
the phase-transition picture, and the contact-order dichotomy
(Section~\ref{sec:contact}). Two diagnostic facts recorded here retroactively
explain the shape the proof had to take. First, the obstruction is genuinely
noncommutative (Remark~\ref{rem:abelian}): the commutative shadow certifies at
bounded degree, so the growth lives in the uncontrolled Bob-side degree of the
noncommutative Fej\'er--Riesz factorization --- and indeed the proof's witness
is a \emph{signed} functional that no Hilbert-space state and no smooth
curve of quantum models can realize (in a Krein space the argument of
\cite{proof} forces the gain direction to be neutral, and no claim is
made). Second, the asymmetric
Alice-degree-one truncations, natural to the Fej\'er--Riesz picture, are the
wrong object ($\mathrm{Asym}(1,m)$ is a sub-basis of the symmetric level
$m{+}1$, so its overshoot only \emph{upper}-bounds $a_{m+1}$); the proof
accordingly works with a positive certificate at the symmetric level itself,
against the right-hand curvature at the convex kink ($c_k''(0^-)=0$,
$c_k''(0^+)=2a_k$) that no convexity or one-sided-Taylor argument alone could
force positive. What remains open after \cite{proof} is quantitative, not
qualitative: the decay law of the sequence $a_k\downarrow0$ (the proof gives
positivity with witness-dependent constants, not rates), and the sharp
dimension law of the first-order space.

\end{document}